\documentclass[final,3p,twocolumn]{elsarticle}

\usepackage{lineno,hyperref}
\usepackage{amsmath}
\usepackage{graphicx}
\usepackage{epstopdf}
\usepackage{float}
\usepackage{multirow}
\usepackage{hhline}
\usepackage{xcolor}

\makeatletter
\newcommand{\vast}{\bBigg@{4}}
\newcommand{\Vast}{\bBigg@{5}}

\def\ps@pprintTitle{%
   \let\@oddhead\@empty
   \let\@evenhead\@empty
   \let\@oddfoot\@empty
   \let\@evenfoot\@oddfoot
}

\makeatother

\begin{document}

\begin{frontmatter}

\title{FIER: Software for analytical modeling of delayed gamma-ray spectra}

\author[1]{E.F. Matthews\corref{cor1}}
\ead{efmatthews@berkeley.edu}
\cortext[cor1]{Corresponding author}

\author[1]{B.L. Goldblum}
\author[1,2]{L.A. Bernstein}
\author[2]{B.J. Quiter}
\author[1]{J.A. Brown}
\author[3]{W. Younes}
\author[3]{J.T. Burke}
\author[3]{S.W. Padgett}
\author[3]{J.J. Ressler}
\author[3]{A.P. Tonchev}

\address[1]{Department of Nuclear Engineering, University of California, Berkeley, California 94720 USA}
\address[2]{Nuclear Science Division, Lawrence Berkeley National Laboratory, Berkeley, California 94720 USA}
\address[3]{Lawrence Livermore National Laboratory, Livermore, California 94550 USA}

\begin{abstract}

A new software package, the Fission Induced Electromagnetic Response (FIER) code, has been developed to analytically predict delayed $\gamma$-ray spectra following fission. FIER uses evaluated nuclear data and solutions to the Bateman equations to calculate the time-dependent populations of fission products and their decay daughters resulting from irradiation of a fissionable isotope. These populations are then used in the calculation of $\gamma$-ray emission rates to obtain the corresponding delayed $\gamma$-ray spectra. FIER output was compared to experimental data obtained by irradiation of a $^{235}$U sample in the Godiva critical assembly. This investigation illuminated discrepancies in the input nuclear data libraries, showcasing the usefulness of FIER as a tool to address nuclear data deficiencies through comparison with experimental data. FIER provides traceability between $\gamma$-ray emissions and their contributing nuclear species, decay chains, and parent fission fragments, yielding a new capability for the nuclear science community. 

\end{abstract}

\begin{keyword}
delayed $\gamma$ rays, radiation simulation, nuclear data, fission products, nuclear forensics
\end{keyword}

\end{frontmatter}


\section{Motivation}
\label{intro}

When a nucleus undergoes fission, the products emit both prompt and delayed characteristic $\gamma$ rays. Delayed $\gamma$-ray emission provides a signature for fissionable materials, with applications in nuclear energy, nuclear forensics, and isotope production. The Fission Induced Electromagnetic Response (FIER) code was developed to meet the need for a transparent, analytical tool to model delayed $\gamma$-ray spectra. Building on the work of Chivers, et al. \cite{chivers}, FIER implements analytical solutions to the Bateman equations, a series of differential equations describing the transmutation of radioactive species \cite{Bateman1910, Pigford1981}. Given a nuclear data library composed of half-lives, decay modes, branching ratios, $\gamma$-ray intensities, and independent fission yields, FIER calculates fission product populations and time-dependent delayed $\gamma$-ray spectra resulting from a user-specified irradiation scheme. 

Simulation of delayed $\gamma$-ray spectra can be performed with other software packages, including MCNP6/CINDER \cite{MCNP, CINDER}, Geant4 \cite{Hecht}, and SCALE/ORIGEN-S \cite{ORIGEN, Williams}. These packages specialize in stochastic simulations that account for geometric effects and particle transport. FIER complements these tools with a 0-dimensional, analytical model that directly links the input nuclear data with the delayed $\gamma$-ray spectral output. Gamma-ray emissions in FIER are quantified as discrete peaks. Thus contributions to a specific energy range can be traced to the relevant discrete $\gamma$ rays, their emitters, the decay chains that produce each emitter, and the originating fission fragments. This inherent transparency in the model output aids the investigation of discrepancies in evaluated nuclear data libraries through comparison with experimental data, as demonstrated in Sec.~\ref{benchmark}. This capability is an important feature of FIER as the need for accurate decay data, $\gamma$-ray intensities, and fission yields has been identified as a critical need of the nuclear science community \cite{NDNCA}. 

A mathematical description of the production and decay of radioactive species is provided in Sec.~\ref{theory}. The computational methodology of FIER is outlined in Sec.~\ref{method}, including a detailed description of input parameters and output data and structure. Model function implementation, uncertainty quantification, options for interfacing with MCNP, and a basic validation of FIER are also presented. A comparison of FIER output with experimental data is given in Sec.~\ref{benchmark}, highlighting the capability of FIER to address nuclear data needs. Concluding remarks are provided in Sec.~\ref{conc}.

\section{Analytical Transmutation Model}
\label{theory}

Gamma rays resulting from fission are classified as prompt or delayed, based on the timescales of the mechanisms governing their emission. Prompt $\gamma$ rays are emitted in the nuclear relaxation of the resultant fission fragments less than a picosecond after the fission event \cite{Peisach1981} and are not simulated by FIER. Delayed $\gamma$ rays, the focus of FIER, are emitted through the population of excited nuclear states that result from the decay of fission fragments and their daughters. As these species have a wide range of half-lives, the emission of delayed $\gamma$ rays occurs on a timescale governed by the decay of fission products and their daughters.

The time-dependent transmutation of the population of a radioactive decay chain is described by a set of coupled differential equations termed the Bateman equations \cite{Bateman1910, Pigford1981}. The rate of change of the population of the first species of a decay chain, $N_{1}(t)$, is given by its production rate less its decay rate:
\begin{equation}
\label{parent}
\frac{\partial N_{1}}{\partial t} = \gamma_{1} R - \lambda_{1} N_{1}(t),
\end{equation}
where $\gamma_{1}$ is the independent fission yield of the parent of the decay chain, $\lambda_1$ is the decay constant of the parent, and $R$ is a constant fission rate. The rate of change of each subsequent species in the decay chain, $N_{i}(t)$, is given by the growth of the species from the activity of its parent less its decay rate:
\begin{equation}
\label{daughter}
\frac{\partial N_{i}}{\partial t} = \beta_{i} \lambda_{i-1} N_{i-1} - \lambda_{i} N_{i}(t),
\end{equation}
where $\lambda_{i}$ is the decay constant of the $i^{th}$ species and $\beta_{i}$ is the branching ratio for decay of the $(i-1)^{th}$ species to the $i^{th}$ species of the decay chain. 

In the case where $R = 0$, an initially pure radioactive sample is allowed to decay. Sequential solution of Eqs.~\ref{parent} and~\ref{daughter} with application of the initial conditions $N_{1}(0) = N_{1}^{0}$ and $N_{i \neq 1}(0) = 0$ yields the population of the $i^{th}$ member of the decay chain:
\begin{equation}
\label{batch}
N_{i}(t) = \Big[ N_{1}^{0} \prod_{l=1}^{i-1} \beta_{l+1}\lambda_{l}\Big] \sum\limits_{j=1}^{i} \frac{ e^{-\lambda_{j}t} }{ \prod\limits_{\substack{k=1 \\ k\neq j}}^{i}(\lambda_{k} - \lambda_{j}) }. 
\end{equation}
Equation~\ref{batch} is referred to as the batch decay solution, as it describes the decay of a ``batch'' of the parent species. 

In the complementary case, where $R \neq 0$, the chain parent is continuously produced by an external process, such as nuclear fission. Sequential solution of this series of differential equations with the initial conditions $N_{i}(0) = 0$ yields \cite{Pigford1981}:
\begin{equation}
\label{cont}
N_{i}(t) = \Big[ \gamma_{1} R \prod_{l=1}^{i-1} \beta_{l+1}\lambda_{l} \Big] \sum\limits_{j=1}^{i} \frac{ 1 - e^{-\lambda_{j}t} }{ \lambda_{j} \prod\limits_{\substack{k =1 \\ k \neq j}}^{i} (\lambda_{k} - \lambda_{j}) }.
\end{equation}
Equation~\ref{cont} is referred to as the continuous production solution; it is used to calculate newly produced populations and their simultaneous decay. Species may be populated by more than one chain parent. This scenario is addressed in FIER by implementing individual decay chains and stems for each species, as described in Sec. \ref{implementation}.

Equation~\ref{cont} becomes numerically unstable when calculating the population of stable or very long-lived species in a decay chain. When the decay constant of a species $j$ is zero (i.e., the species is stable), both the numerator and denominator inside of the summation are zero yielding an indeterminate result. Similarly, if a non-stable species has a very small decay constant then the numerator inside of the summation may appear to be zero due to the finite bit resolution of the computer. To resolve this issue, this species must be treated as stable and any species in the decay chain following it will have no population. To calculate the population of a stable or very long-lived species $n$, the $n^{th}$ term inside of the summation in Eq.~\ref{cont} is expanded using a Taylor series, which evaluates to:
\begin{equation}
\label{stable_cont}
\frac{ 1 - e^{-\lambda_{n}t} }{ \lambda_{n} \prod\limits_{\substack{k =1 \\ k \neq n}}^{n} (\lambda_{k} - \lambda_{n}) } \rightarrow \frac{ t }{ \prod\limits_{\substack{k =1 \\ k \neq n}}^{n} (\lambda_{k} - \lambda_{n}) }  .
\end{equation}

Equations~\ref{batch} and~\ref{cont} are also unstable when two species in a decay chain, $k$ and $j$, have the same decay constant. In this situation the denominator inside of the summation is zero while the numerator is finite, yielding a non-physical, infinite population for species $i$. Solutions to this artifact include: the decay constants of species $j$ and $k$ can be varied slightly within their associated uncertainty to achieve an approximate solution or the reciprocal of the product in the denominator can be expanded using a Taylor series approximation \cite{Cetnar}. The former solution is implemented in FIER.

\section{Methods}
\label{method} 

\subsection{Input Data}

The computational implementation of the Bateman solutions necessitates assembly of the decay chains for each fission fragment and its daughters. This requires the decay modes, associated branching ratios, and half-lives of the fission fragments and their daughters as input data. Using empirical independent fission yields and a user-specified irradiation scheme, FIER calculates the population of fission fragments and their daughters at the end of a specified irradiation. As Eq.~\ref{cont} requires a constant fission rate, the user-specified irradiation scheme is composed of a series of time intervals, each with a corresponding constant fission rate. Continuously time-varying irradiation schemes can be simulated by discretizing the irradiation using a series of constant fission rates. FIER does not account for burn-up of the fissioning material, so if depletion is a concern, the user must decrease the fission rate with time accordingly. The calculation of the resultant time-dependent delayed $\gamma$-ray spectra requires the $\gamma$-ray energies and intensities for each species as input data. The user specifies a counting scheme composed of the desired time intervals to obtain the resultant delayed $\gamma$-ray spectra. For each of these time intervals, FIER calculates and outputs the expected number of $\gamma$ rays emitted. 

The sources for the input dataset used in this work are summarized in Table \ref{data}. The source of independent fission yields was the England and Rider library \cite{Rider1994}. The $\gamma$-ray energies and intensities, decay modes and branching ratios, and half-lives were taken from the ENDF Decay Sub-Library \cite{ENDFDecay}. For the experimental study presented in Sec.~\ref{benchmark}, the data from the ENDF Decay Sub-Library were inspected for accuracy against the community standard for structure and decay data: ENSDF \cite{ENSDF} . For the study in Sec.~\ref{benchmark}, the FIER input data library was updated to use the ENSDF values in those cases where the ENDF and ENSDF library values disagreed.

\begin{table*}
\center
\begin{tabular}{ l  l }
\hline
Parameters & Source(s) \\
\hline
Irradiation Scheme & User Specified \\ 
Counting Scheme & User Specified \\
Fission Yields & England and Rider \cite{Rider1994} \\
$\gamma$ Ray Energies and Intensities & ENDF Decay Sub-Library \cite{ENDFDecay}, ENSDF \cite{ENSDF} \\
Decay Modes and Branching Ratios & ENDF Decay Sub-Library \cite{ENDFDecay}, ENSDF \cite{ENSDF} \\
Half-lives/Decay Constants & ENDF Decay Sub-Library \cite{ENDFDecay} \\
\hline
\end{tabular}
\caption{Sources of nuclear data for simulations presented in this study.}
\label{data}
\end{table*}

\subsection{Implementation of the Bateman Solutions}
\label{implementation}

At initialization, FIER generates a list of all possible decay chains for each fission fragment and its daughters. As radioactive species can have multiple decay modes, a given species can appear in multiple decay chains. Using these decay chains, FIER identifies all decay paths that lead to each species, termed \textit{decay stems}. For example, consider this representative radioactive decay chain:
\begin{equation*}
^{132}\text{Te} \rightarrow {^{132}\text{I}} \rightarrow {^{132}\text{Xe}}
\end{equation*}
It contains the following three decay stems:
\begin{align*}
^{132}\text{Te} &  \\
^{132}\text{Te} & \rightarrow {^{132}\text{I}}  \\
^{132}\text{Te} & \rightarrow {^{132}\text{I}} \rightarrow {^{132}\text{Xe}}
\end{align*}

Each species has a set of decay stems associated with it and the total population of a species at time $t$ is the sum of the populations calculated for each of its associated decay stems. For each decay stem, the population of its associated species (i.e. the last species in the stem) is calculated using Eqs. \ref{batch} and \ref{cont}. At $t=0$, the initial population of a species is assumed to be zero if no initial population is specified by the user. To avoid double counting, only the last species in a stem is counted. Performing the calculations using decay stems allows the otherwise complex and highly coupled system of decay chains to be linearized. For example, the radioactive species $^{132}$Sn is fed by the branched decay of multiple parents. To illustrate how decay stems are used to linearize decay chains, the decay stems of $^{132}$Sn are shown:
\begin{align*}
^{132}\text{Sn} & \\
^{132}\text{In} & \rightarrow {^{132}\text{Sn}} \\
^{133}\text{In} & \rightarrow {^{132}\text{Sn}} \\
^{132}\text{Cd} & \rightarrow {^{132}\text{In}} \rightarrow {^{132}\text{Sn}} \\
^{133}\text{Cd} & \rightarrow {^{133}\text{In}} \rightarrow {^{132}\text{Sn}}
\end{align*}

\begin{figure}
\center
\includegraphics[width=0.5\textwidth]{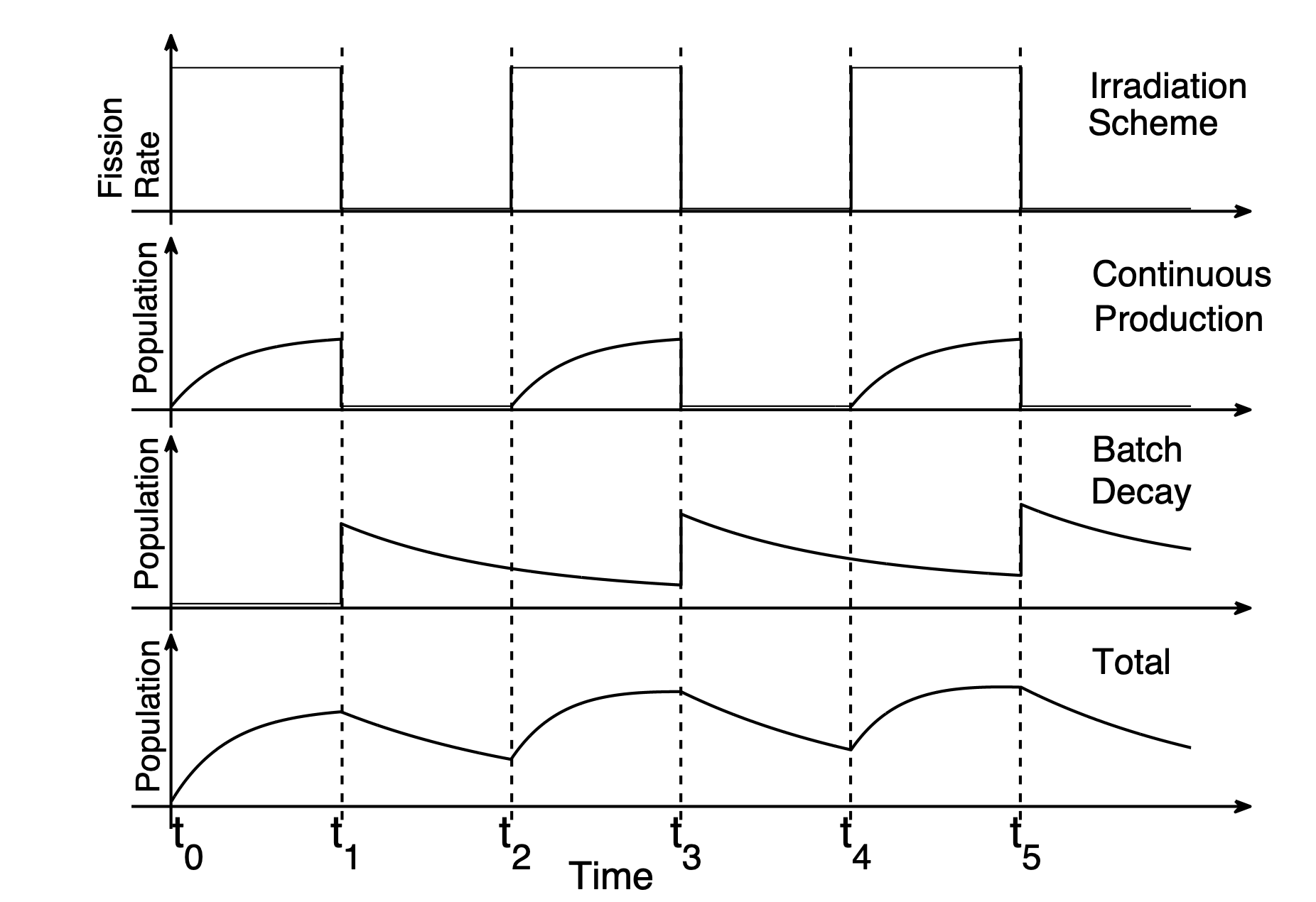}
\caption{Illustration of the implementation of the Bateman solutions. The population calculated in the continuous production solution at the end of each time step is added to the initial population in the batch decay solution for the subsequent time step. \label{scheme}}
\end{figure}

An example of the computational implementation of the continuous production and batch decay solutions (i.e., Eqs. \ref{batch} and \ref{cont}) is illustrated in Fig.~\ref{scheme} for a parent species. As it is not fed by the decay of other species, a parent species will only grow and decay by single exponentials. With an initial population of zero, the population of the species produced over the time interval $t_{0}$ to $t_{1}$ is calculated using the fission rate specified by the irradiation scheme and Eq. \ref{cont}. For the next time interval spanning $t_{1}$ to $t_{2}$, the total population at time $t_{1}$ is used as the initial population in the batch decay solution given by Eq. \ref{batch}. In this illustration, there is no fission during this time interval, and thus the total population at time $t_{2}$ is the result of the batch decay solution only. 

For the next time interval spanning $t_{2}$ to $t_{3}$, the population of the species at time $t_{2}$ is used as the initial population in Eq. \ref{batch}. Eq. \ref{cont} is used to calculate the production and simultaneous decay of new species based on the fission rate specified by the irradiation scheme for this time interval. These two populations are summed to give the total population at the end of the time interval, $t_{3}$. This process continues for each time interval specified in the irradiation scheme and the batch decay solution is used to calculate populations for any time specified after irradiation has ceased. 

\subsection{Calculation of Delayed $\gamma$-Ray Spectra}

The counting scheme begins following irradiation and any user-specified cooling period. During the cooling and counting period, the decay of the product populations is calculated using Eq.~\ref{batch}. To calculate the number of $\gamma$-ray emissions as a function of time, a user-specified set of time intervals is employed. For a time interval, $t_{n}$ to $t_{n+1}$, the number of emissions, $C_{i}^{m}$, of the $m^{th}$ $\gamma$ ray of intensity, $I_m$, from the $i^{th}$ species of the decay chain is the integral of the time-dependent $\gamma$-ray emission rate:
\begin{equation}
C_{i}^{m} = \int\limits_{t_{n}}^{t_{n+1}} I_{m} \lambda_{i} N_{i}(t) dt.
\end{equation}
Substituting the population given by the batch decay solution in Eq. \ref{batch} and integrating yields:
\begin{equation}
\label{batch_counting}
C_{i}^{m} = I_{m} \lambda_{i} N_{1}^{0} \Big[ \prod_{l=1}^{i-1} \beta_{l+1}\lambda_{l} \Big]     \sum\limits_{j=1}^{i} \frac{ -e^{-\lambda_{j}t} }{ \lambda_{j} \prod\limits_{\substack{k=1 \\ k\neq j}}^{i}(\lambda_{k} - \lambda_{j}) } \Bigg|_{t_{n}}^{t_{n+1}}. 
\end{equation}

\subsection{Design and Output}
\label{output_sec}

\begin{figure}
\center
\includegraphics[width=0.5\textwidth]{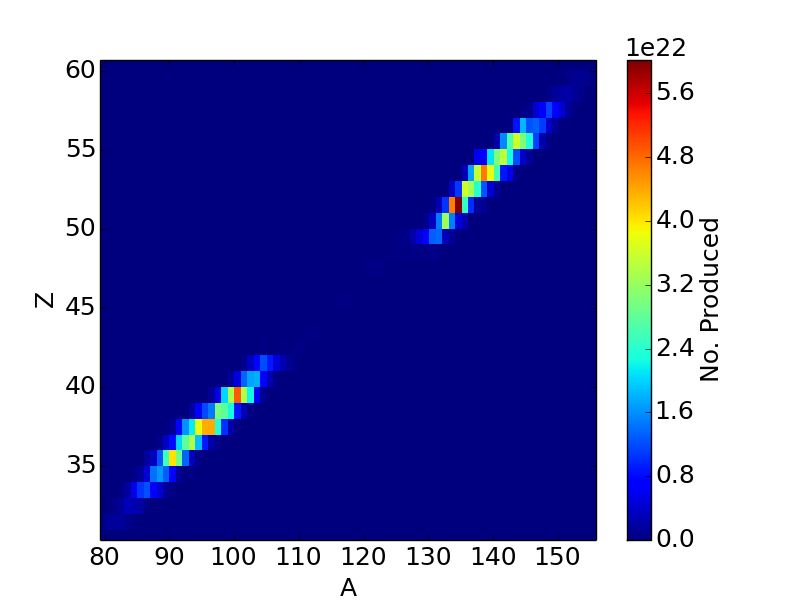}
\caption{Population distribution in integer bins of $^{235}$U fission products after 1 $\mu$s of fission-spectrum neutron irradiation resulting in 10$^{24}$ fissions. Note the zero-suppressed axes. There are non-zero populations for species with $Z$ ranging from 23 to 70. However, only the region displayed has populations large enough to be seen on a linear color map.}
\label{pops}
\end{figure}

\begin{figure}
\center
\includegraphics[width=0.5\textwidth]{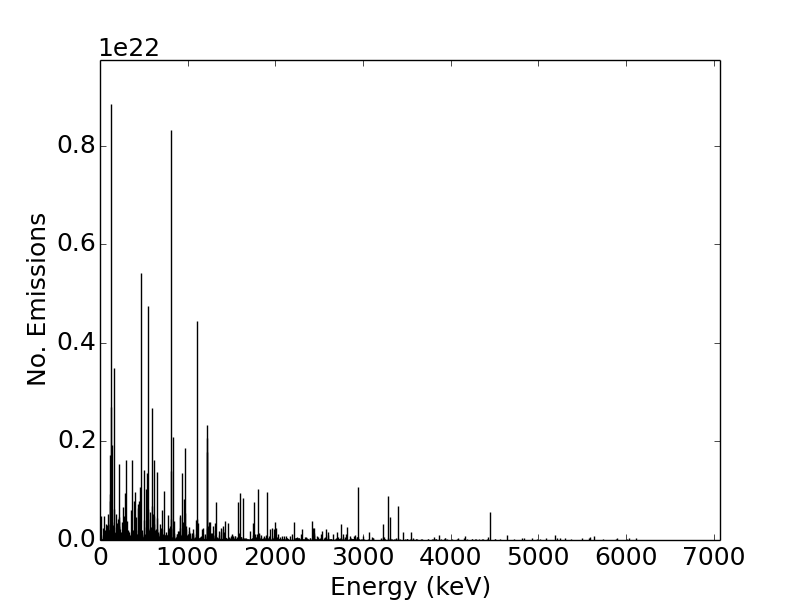}
\caption{Discrete line plot of the delayed $\gamma$-ray spectrum emitted between 1 and 2 s after a 1~$\mu$s fission-spectrum neutron irradiation of isotopically pure $^{235}$U resulting in 10$^{24}$ fissions.}
\label{hist}
\end{figure}

FIER is a standalone C++ program with no external packages or dependencies. The FIER output includes both fission product population data and delayed $\gamma$-ray spectra in comma-separated-value format. The former can be obtained for any specified point in time during or after irradiation and the latter for any specified time interval after the irradiation. In the output file, each column corresponds to the time-dependent evolution of the population of a radioactive species and/or the time-dependent evolution of a $\gamma$ ray emitted from a radioactive species. Header rows specify the Z, A, excitation energy, and half-life for the species in each column and/or the emitter Z, A, excitation energy, half-life, and the emitted $\gamma$-ray energy, respectively. Each subsequent row corresponds to a user-specified point in time. For mixed composition samples, the radionuclide populations and delayed $\gamma$-ray spectra can be obtained through the fission-cross-section-weighted superposition of population and spectral data for each component of the mixture. Figures \ref{pops} and \ref{hist} show visualizations of a representative FIER output for a 1 $\mu$s fission-spectrum neutron irradiation of an isotopically pure $^{235}$U sample resulting in 10$^{24}$ fissions.

\begin{figure}
\center
\includegraphics[width=0.45\textwidth]{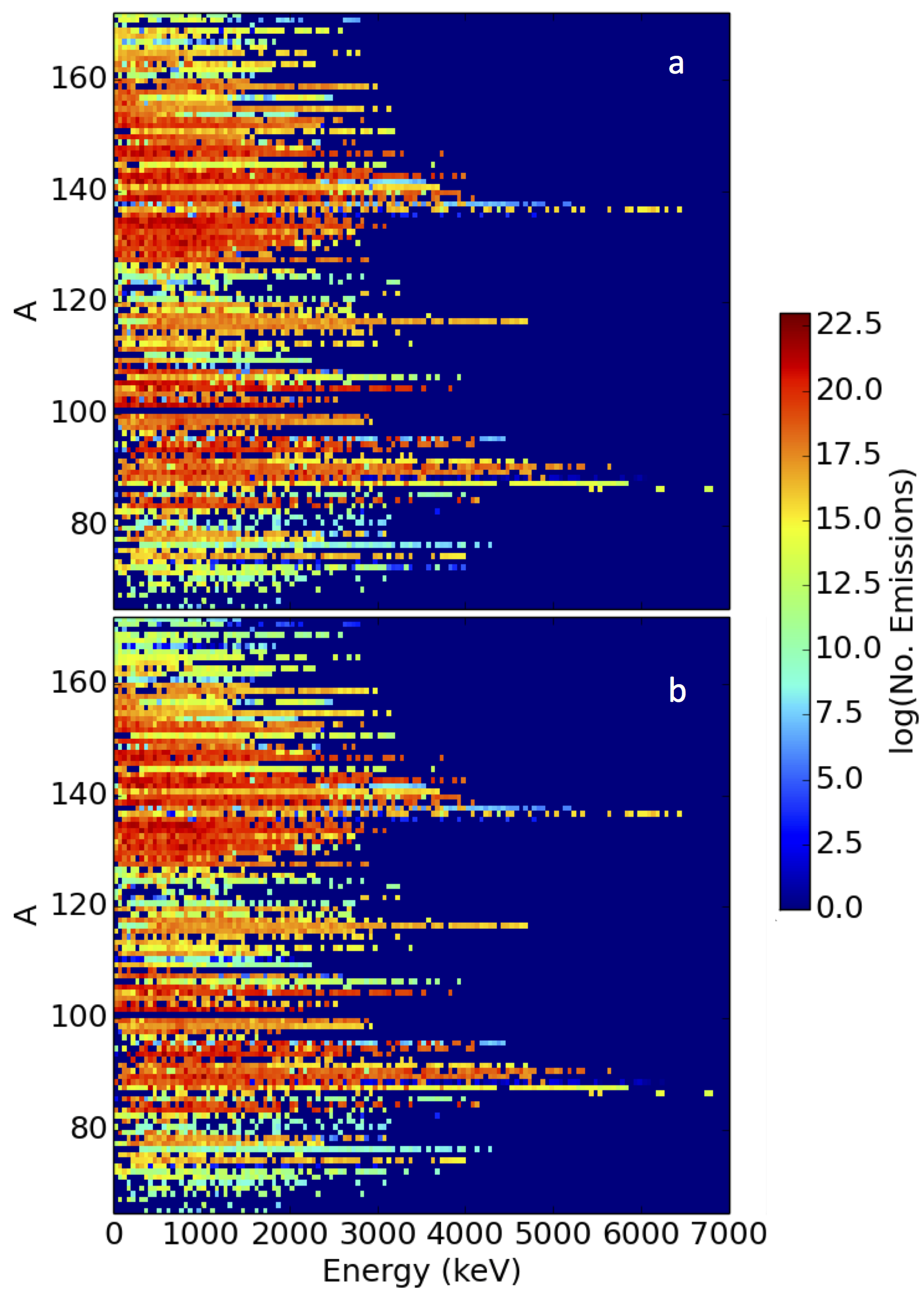}
\caption{ $\gamma$-ray emissions in 50.0 keV energy bins over the time interval 1000-2000 s after a 1 $\mu$s of fission-spectrum neutron irradiation of isotopically pure a) $^{235}$U and b) $^{239}$Pu. Although difficult to see on a log-scale color map, there are significant differences between these spectra shown in Fig.~\ref{fissile_diff}. }
\label{combined_1e3}
\end{figure}

\begin{figure}
\center
\includegraphics[width=0.5\textwidth]{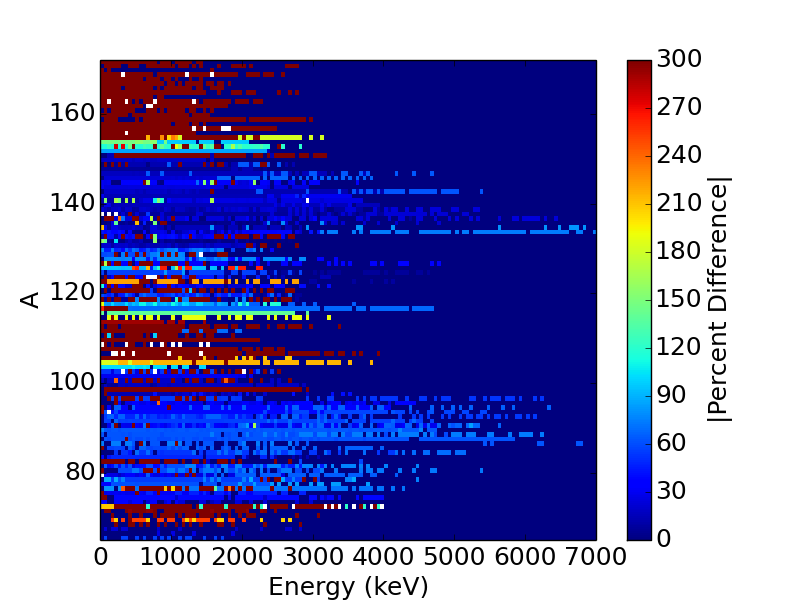}
\caption{ Absolute value of the percent difference between the binned spectra of $^{235}$U and $^{239}$Pu shown in Figs.~\ref{combined_1e3}a and \ref{combined_1e3}b. White pixels represent regions where $^{235}$U had no emissions but $^{239}$Pu did. Dark red represents regions with percent difference greater than or equal to $\pm$300$\%$. }
\label{fissile_diff}
\end{figure}

The $\gamma$-ray emissions over the time interval 1000-2000 s after 1 $\mu$s of fission-spectrum neutron irradiation of $^{235}$U and $^{239}$Pu are shown in 50.0 keV energy bins in Figs.~\ref{combined_1e3}a and \ref{combined_1e3}b, respectively. These results showcase the flexibility the user has over FIER output in that $\gamma$-ray emissions can be investigated as a function of product mass number. The results in Fig.~\ref{combined_1e3} are also suggestive of the nuclear forensics applications of FIER. Specifically, there are significant differences in the delayed $\gamma$-ray emissions resulting from identical irradiations of different fissile materials. The absolute value of the percent difference between the binned spectra in Fig.~\ref{combined_1e3} is shown in Fig.~\ref{fissile_diff}. The mass regions associated with the heavy and light fragment wings and the symmetric fission valley of the fission yield distribution present regions with significantly different emissions, owing to the differences in product yields between $^{235}$U and $^{239}$Pu. These differences offer key information that can be used to discriminate between samples with different fissile material compositions. 

Additionally, FIER can output a list of possible decay chains and stems for each fission product. Thus, population and $\gamma$-ray emission data can be traced back to the decay chains/stems and fission products that contribute to them. The delayed $\gamma$-ray spectra simulated by FIER can also be used to generate source cards for use in Monte Carlo N-Particle (MCNP) simulations. This allows FIER outputs to be transported through a material geometry, providing the user with both transported and original source delayed $\gamma$-ray spectra. This capability can offer powerful insight into experimentally observed spectra, assisting nuclear power and forensics applications. 

\subsection{Uncertainty Quantification}
\label{unc_analysis}

FIER utilizes empirical nuclear data as input and the associated uncertainty on these nuclear data are propagated in the FIER model. A Monte Carlo method is used as a probabilistic approach to uncertainty quantification. This approach has been coded as an optional feature of FIER. The input nuclear data (i.e., the $\gamma$ ray intensities and energies, branching ratios, decay constants, and fission yields) are statistically varied about a Gaussian distribution with a mean of the nominal value and a standard deviation of the uncertainty on the parameter. The FIER calculations are repeated for each randomly sampled input library to produce a distribution of possible outcomes. The standard deviation in the model output is then calculated from the standard deviation in the various trial results. This allows for an assessment of the impact of the uncertainty associated with individual nuclear data parameters on the resulting population and spectral data. Results from uncertainty quantification in FIER are showcased in Sec.~\ref{benchmark}.

\subsection{Model Validation}
The population calculations were validated by comparing FIER calculated chain yields to evaluated chain yields from England and Rider \cite{Rider1994}. To maintain consistency, the beta-neutron emission branching ratios in England and Rider were used as input to FIER.  Using thermal $^{235}$U independent fission yields and a very short irradiation period of 10 $\mu$s at 10$^{7}$ fissions per second, 200 fission products were produced. The transmutation of these nuclides was then assessed at 10$^{8}$ seconds after irradiation, long after beta-neutron emission ceased. The chain yields were then calculated by summing the FIER-calculated populations for all species of a given mass number. These were compared to the England and Rider evaluated chain yields and all 107 chain yields (A = 66-172) agreed with the England and Rider evaluation to within 2$\%$, with 97 chain yields agreeing to within 0.25$\%$. Moreover, all of the chain yields agreed well within the uncertainties listed in England and Rider. 

The most discrepant chain yields were mass chains either in the light fragment wing or the symmetric fission valley of the fission yield distribution. This is expected as both of these regions have low yields with large relative uncertainties. The small discrepancies between the FIER chain yields and the England and Rider chain yields are suspected to arise from the evaluation process used by England and Rider, specifically the variance-weighted normalization method used to ensure the measured chain yields summed to 200$\%$ \cite{Rider1994}.

\section{Comparison with Experimental Data}
\label{benchmark}

Output from FIER was compared to delayed $\gamma$-ray spectra from a 63-mg highly-enriched $^{235}$U sample ($>$99 wt$\%$) irradiated at the Godiva critical assembly at the National Criticality Experiments Research Center at the Nevada National Security Site \cite{Stave}. The emission data were collected using two 60$\%$ efficient (at 1.33 MeV relative to a 3"$\times$3" NaI detector) Canberra model BE6530 high-purity Germanium (HPGe) detectors. The neutron irradiation resulting from the Godiva assembly was a very short pulse on the order of tens of microseconds. Counting of the sample commenced 291 minutes following irradiation and continued for one week. Data from each HPGe detector were collected in time-stamped, list-mode format with an event clock timing resolution of 100~ns. 

The $\gamma$-ray yield data were extracted from the experimental spectra using the XGAM spectrum-fitting code \cite{Walid2014}. These measured $\gamma$-ray yields were corrected using the absolute detection efficiency of the HPGe detectors and the data acquisition dead-time to obtain the number of emissions. The absolute detection efficiencies were obtained using well characterized $^{22}$Na, $^{60}$Co, $^{137}$Cs, $^{133}$Ba, and $^{152}$Eu calibration sources. The $\gamma$ emissions of these calibration sources spanned the energy range from 50 keV to 1.4 MeV. The spectra from these calibration sources were used to determine an energy-dependent efficiency for the detector of the functional form:
\begin{equation}
\label{abs_eff}
\varepsilon(E_{\gamma}) = P_{1}\,E_{\gamma}^{-P_{2}} + P_{3} - P_{4} e^{ -P_{5}\, E_{\gamma} } ,
\end{equation}
where $\varepsilon(E_{\gamma})$ is the absolute detection efficiency for a $\gamma$ ray of energy $E_{\gamma}$ and $P_{1}$ - $P_{5}$ are fitting parameters. The uncertainty of the detector efficiency was on the order of a few percent throughout the entire energy range studied and these uncertainties were propagated in the XGAM analysis. 

FIER was used to simulate the delayed $\gamma$-ray spectra resulting from fission of $^{235}$U using the England and Rider fission-spectrum neutron independent yield library \cite{Rider1994} and an irradiation scheme matching that of the experiment. The number of fissions in the experiment was determined to be 5.3$\times$10$^{10}$ using a witness foil and the FIER output was normalized to match this \cite{Walid2014}. A total of 163 1-hour spectra were collected, from which 564 photopeaks were identified. The counts in each photopeak were analyzed, efficiency corrected, and assigned to the emitting species using a $\gamma$-ray energy and half-life matching technique. These data were then compared to the FIER model output using the following figure-of-merit, $\varepsilon$:
\begin{equation}
\label{merit}
\varepsilon = 100 \frac{Y_{o} - Y_{e}}{Y_{e}},
\end{equation}
where $Y_{o}$ is the observed $\gamma$-ray yield, obtained from the experimental data, and $Y_{e}$ is the expected $\gamma$-ray yield predicted by FIER. 

\begin{figure}
\center
\includegraphics[width=0.5\textwidth]{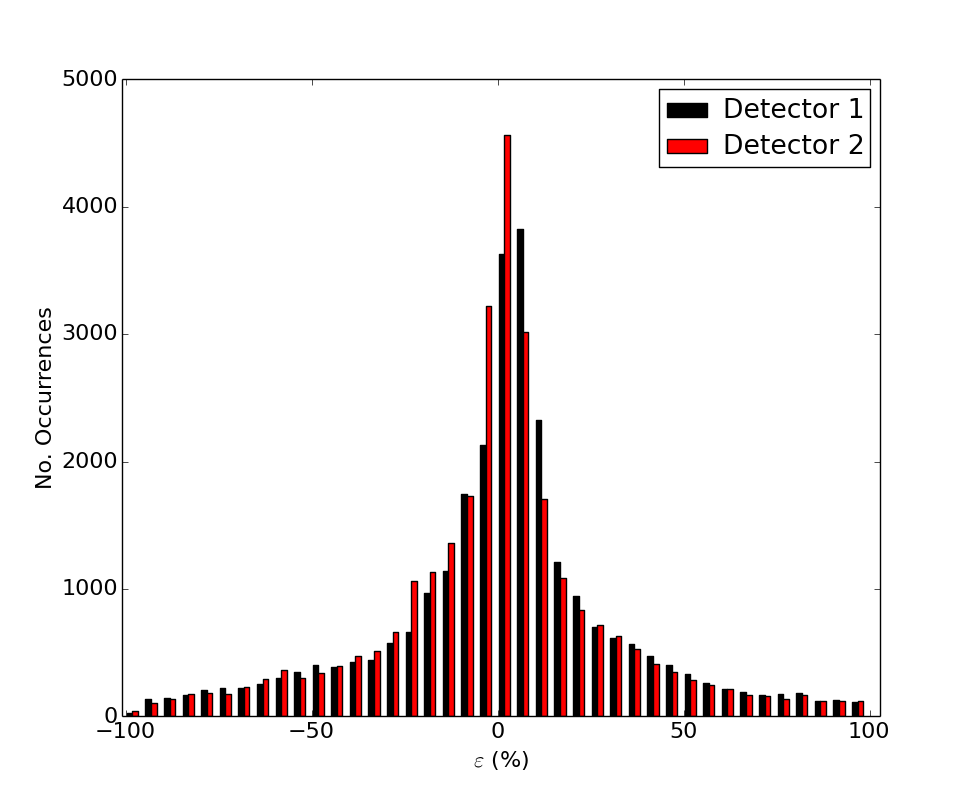}
\caption{Comparison of FIER simulation to experimental data \cite{Walid2014}. Plotted is a histogram with $5\%$ bins of the figure-of-merit defined by Eq. \ref{merit} calculated for each $\gamma$ ray observed. Detectors 1 and 2 represent the numerical labels of the two detectors used in the collection of the experimental data.}
\label{comp}
\end{figure}

The figure-of-merit for each observed photopeak is presented in a time-integrated histogram in Fig.~\ref{comp}. The results show fair agreement with the mean value of $\varepsilon$ being -0.48$\%$ for Detector 1 and -2.36$\%$ for Detector 2. However, it can be seen there is still significant disagreement between the FIER model output and the experimental yields; the distribution for the two detectors have standard deviations of 32.46$\%$ and 31.33$\%$, respectively. Disagreements between FIER and the experimental data can be attributed to two possible sources: uncertainties in the $\gamma$-ray yields related to the experimental and/or analysis methods and discrepancies in the evaluated nuclear data libraries. As the model implemented in FIER uses evaluated nuclear data as input, the accuracy of the FIER output is dependent on the accuracy of the input nuclear data. In particular, some of the independent fission yields input to FIER have substantial uncertainties. The average relative uncertainty in the independent fission yields for the England and Rider $^{235}$U fission-spectrum neutron library is 59.9$\%$. This uncertainty is reflected in the distributions shown in Fig.~\ref{comp}, where 90$\%$ of the data for the distributions of both detectors fall within bins $\pm55\%$ from their respective mean values of $\varepsilon$.

\begin{figure}
\center
\includegraphics[width=0.5\textwidth]{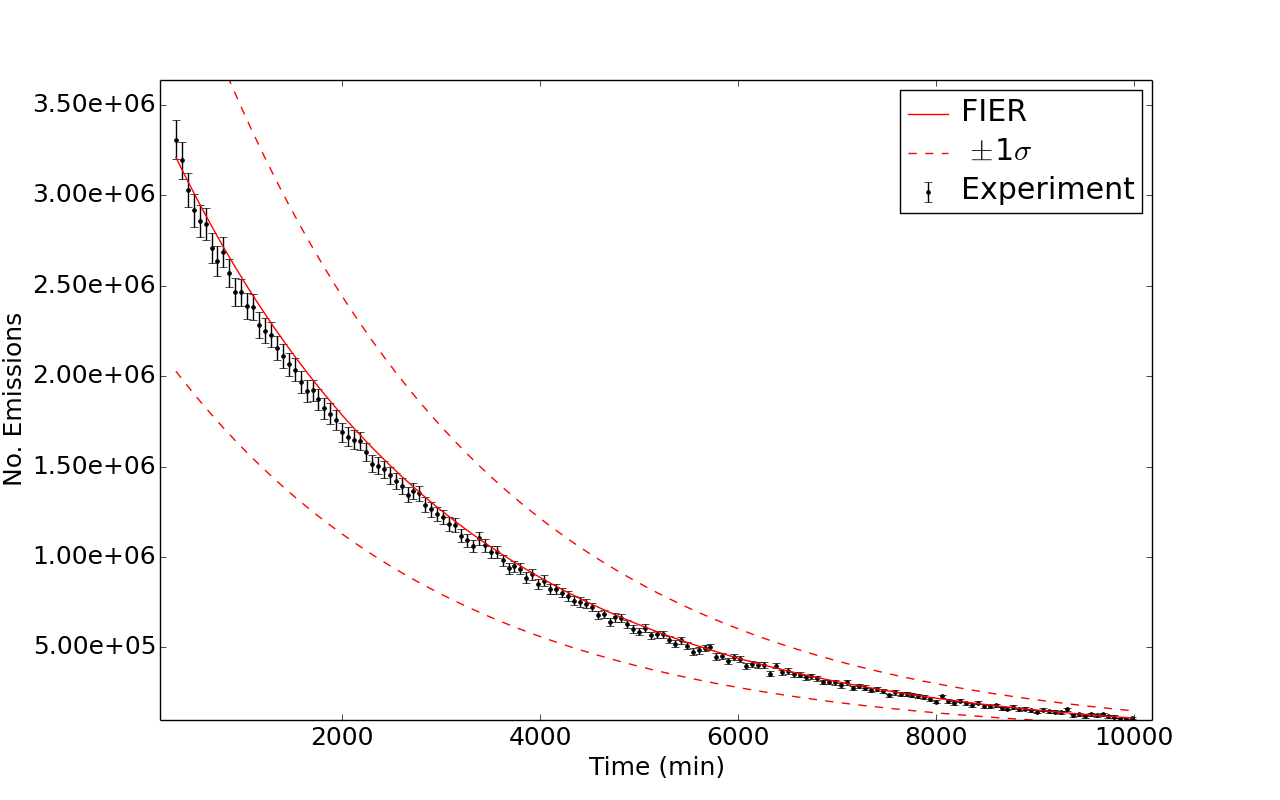}
\caption{Time-dependent evolution of the $664.6$ keV $\gamma$-ray emission from the decay of $^{143}$Ce in 1 hour time intervals starting 291 minutes after the end of irradiation. The dashed lines represent one standard deviation uncertainty bounds in the FIER model output.}
\label{good}
\end{figure}

\begin{figure}
\center
\includegraphics[width=0.5\textwidth]{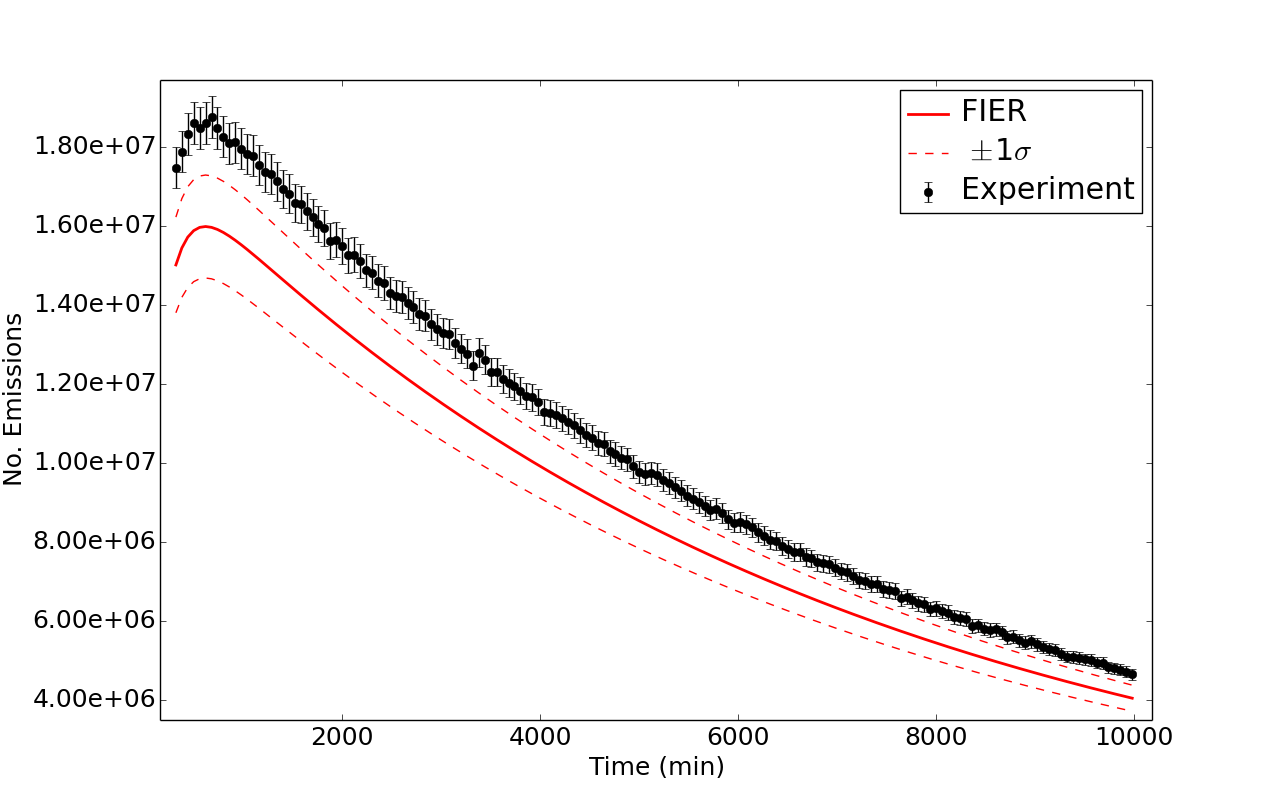}
\caption{Time-dependent evolution of the $667.7$ keV $\gamma$-ray emission from the decays of $^{132}$I, $^{132m}$I, $^{132}$Cs, and $^{127}$Sb in 1 hour time intervals starting 291 minutes after the end of irradiation. The dashed lines represent one standard deviation uncertainty bounds in the FIER model output.}
\label{bad1}
\end{figure}

Figures~\ref{good} and \ref{bad1} illustrate comparisons between the FIER output and observed individual $\gamma$-ray transitions with energies of 664.6 keV and 667.7 keV, respectively. Figure~\ref{good} shows the time evolution of the $664.6$~keV $\gamma$ ray emitted in the decay of $^{143}$Ce. The FIER model uncertainty was obtained using the Monte Carlo method described in Sec.~\ref{unc_analysis} with 3000 trials to provide a statistical uncertainty of less than 2$\%$. The FIER output exhibits good agreement with the experimental data over the full time range investigated, suggesting that the input nuclear data associated with this $\gamma$-ray emission are accurate. Figure~\ref{bad1} shows the time evolution of the $667.7$ keV $\gamma$ rays emitted in the decays of $^{132}$I, $^{132m}$I, $^{132}$Cs, and $^{127}$Sb after the end of irradiation. Here, the FIER output is systemically lower than the experimental data throughout the observed time window. 

\begin{figure}
\center
\includegraphics[width=0.5\textwidth]{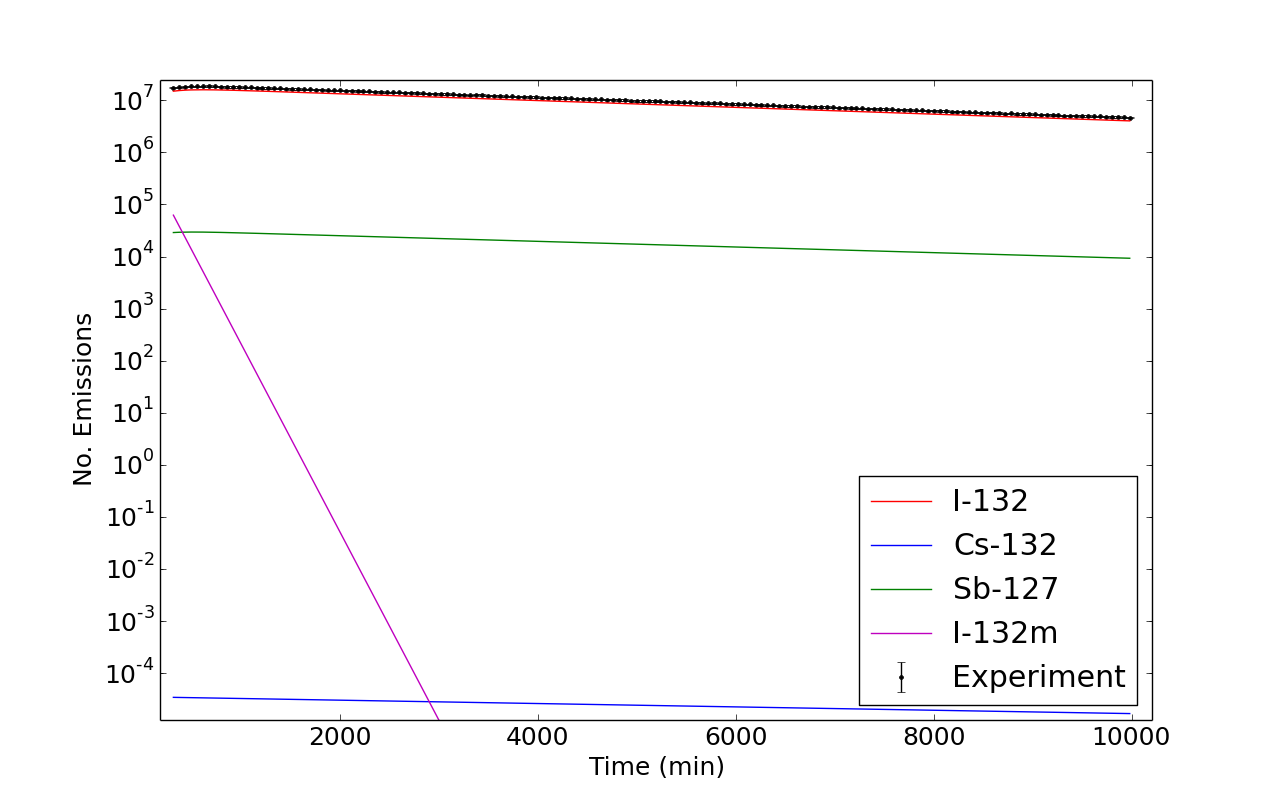}
\caption{Time-dependent evolution of the $667.7$ keV $\gamma$-ray emission from the decays of $^{132}$I, $^{132m}$I, $^{132}$Cs, and $^{127}$Sb, plotted in log scale in 1 hour time intervals starting 291 minutes after the end of irradiation. The contributions from $^{132m}$I, $^{132}$Cs, and $^{127}$Sb are orders of magnitude smaller than that of $^{132}$I.}
\label{bad1_contribs}
\end{figure}

The disagreement between the experimental data and the FIER output shown in Fig.~\ref{bad1} could be caused by inaccuracies in the fission yields of the chain species or by inaccuracies in the $\gamma$-ray intensity and branching ratios. It is also possible that the discrepancy is due to changes in the incident neutron spectrum when comparing the empirical and evaluated fission yields. The results in Fig.~\ref{comp} suggest that the incident neutron spectrum is reasonably well-described by the England and Rider 500-keV fission spectrum, though the shoulders of the double-humped fission product charge distribution are more sensitive to changes in the incident neutron spectrum. Inspection of the relative contributions to this emission by each of the species was investigated as shown in Fig.~\ref{bad1_contribs}, revealing that $^{132}$I dominates the emission by several orders of magnitude. Thus, the discrepancy is driven by nuclear data discrepancies in $^{132}$I or its progenitors. A discrepancy in the $\gamma$-ray intensity of the 667.7 keV transition from $^{132}$I is not suspected to be the source of the disagreement as the $\gamma$-ray intensity is characterized with an uncertainty of less than 1$\%$ \cite{ENSDF}. Likewise, the branching ratios associated with the species in the decay chains leading to $^{132}$I all have either nominal uncertainty or belong to species with very small yield and thus do not contribute strongly to the propagated uncertainty.

In contrast, analysis of the decay stems leading to $^{132}$I shows that the independent fission yields of $^{132}$I and its parents all have relatively large uncertainties ($>$10$\%$). Only the time-dependent decay contributions from $^{132}$I and its parents $^{132}$Te and $^{132m}$I are observable, as only these species have half-lives long enough to be observed after the 291 minute cooling period. Examining the uncertainties on the cumulative yields of these species gives insight into the possible sources of the discrepancy. The cumulative yield of $^{132}$Te is characterized with an uncertainty of 2$\%$ \cite{Rider1994} and the cumulative yield of $^{132m}$I is relatively small; thus neither of these are expected to be the source of the discrepancy. The cumulative yield of $^{132}$I is large and has a relative uncertainty of 64$\%$ \cite{Rider1994} and is therefore suspected to be the source of the discrepancy. This suggests that the disagreement between the FIER output and the experimental data is due to the evaluated fission yields associated with $^{132}$I and its parents. If delayed $\gamma$-ray spectra had been collected sooner after irradiation, it is possible specific independent fission yields could have been identified as the source of the discrepancy. This exercise showcases the use of FIER in identifying nuclear data needs for applications. 

\begin{figure}
\center
\includegraphics[width=0.5\textwidth]{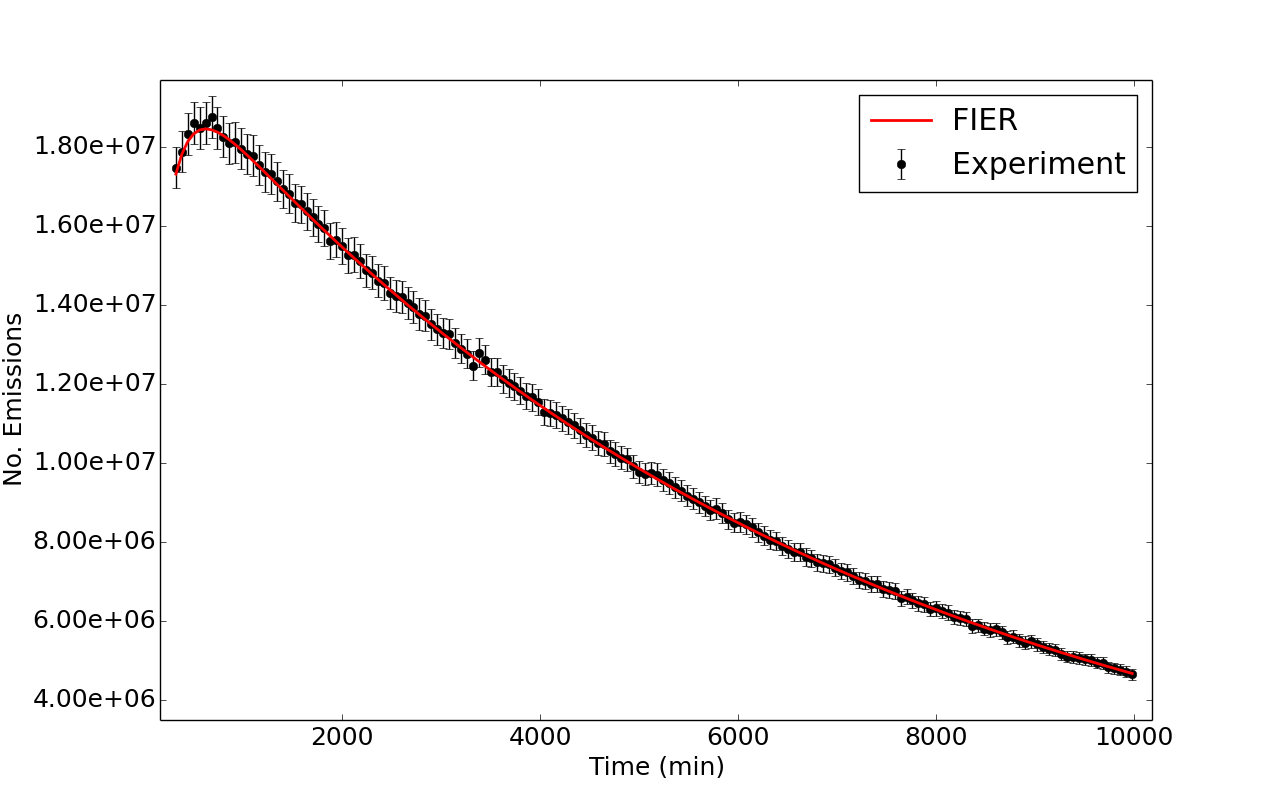}
\caption{FIER model output after the cumulative yield of $^{132}$I was increased by 0.24$\sigma$ to 5.39$\%$. This modest adjustment resolves the disagreement between the experimental data and FIER shown in Fig.~\ref{bad1}.}
\label{after1}
\end{figure}

A solution to this nuclear data discrepancy was investigated by performing a chi-squared minimization and varying the independent yield of a short-lived precursor species of $^{132}$I, which is reflected in the model output as an effective adjustment to the 4.67$\pm$2.99$\%$ \cite{Rider1994} cumulative fission yield of $^{132}$I. It was found that by increasing the cumulative yield of $^{132}$I by 0.72$\%$ (+0.24$\sigma$) to 5.39$\%$ the discrepancy between the experimental data and the FIER model output was resolved. This is shown in Fig.~\ref{after1} and exemplifies how FIER can be used to inform nuclear data evaluations in addition to identifying data discrepancies.

\section{Conclusions}
\label{conc}

Using the Bateman solutions, evaluated nuclear data, and a specified irradiation scheme, FIER produces radioactive decay chains, time-dependent fission product population data, and delayed $\gamma$-ray spectra following fission. FIER has the potential to assist in the improvement of evaluated decay and fission product yield data libraries. By identifying discrepancies between experimental data and FIER model output, experimental activities can be directed to target various nuclear data properties, including half-lives, decay mode branching ratios, $\gamma$-ray intensities, and fission yields \cite{NDNCA}. Beyond basic science applications, FIER can be used to perform calculations in support of various nuclear applications. Further, FIER provides a new capability to improve fission theory and modeling by providing a bridge between delayed gamma-ray measurements and independent fission product yields. Though FIER does not offer inherent transport and depletion capabilities, it delivers a feature not available in existing transmutation code packages by specifying the decay chains and nuclear data associated with each individual $\gamma$-ray transition in the output. The capabilities introduced by FIER offer a new tool for both the applied and basic nuclear science communities.

\section*{Acknowledgements}

The authors would like to thank Larry Greenwood, Amanda Prinke, and Sean Stave of Pacific Northwest National Laboratory for their contributions to the experimental portion of this study. The authors would also like to recognize Isaac C. Meyer and Nidhi R. Patel for early contributions to the research. 

This material is based upon work supported by the Department of Energy National Nuclear Security Administration through the Nuclear Science and Security Consortium under Award Numbers DE-NA0003180 and DE-NA0000979, and performed under the auspices of the U.S. Department of Energy by Lawrence National Security, LLC, Lawrence Livermore National Laboratory under Contract DE-AC52-07NA27344. This work is also supported by the Lawrence Berkeley National Laboratory under Contract No. DE-AC02-05CH11231 for the US Nuclear Data Program. 


\bibliography{bibfile3}

\end{document}